\documentclass[12pt]{article}
\begin{document}
\title{Comment on ``A note on the construction of
the Ermakov-Lewis invariant''}
\author{F.~Haas  \\
Laborat\'orio Nacional de Computa\c{c}\~ao Cient\'{\i}fica \\
Coordena\c{c}\~ao de Matem\'atica Aplicada  \\
Av. Get\'ulio Vargas, 333 \\
25651-070, Petr\'opolis, RJ - Brazil\\
ferhaas@lncc.br\\ \strut\\
J.~Goedert\\
Centro de Ci\^encias Exatas e Tecnol\'ogicas - UNISINOS\\
Av. Unisinos, 950\\
93022-000, S\~ao Leopoldo, RS - Brazil\\ goedert@exatas.unisinos.br}
\date{\relax}
\maketitle
\begin{abstract} \noindent
We show that the basic results on the paper referred in the title [J. Phys. A: Math. Gen. {\bf 35}
(2002) 5333-5345],
concerning the derivation of the Ermakov invariant from Noether
symmetry methods, are not new.
\end{abstract}


\vskip 1cm

\noindent The purpose of this comment is to point out that the
main results presented in a recently published paper \cite{Moyo},
are not new. At the end of the introduction of this paper, the
authors claims that ``\dots {\it this is the first time the
Noether symmetries are being considered to discuss the source of
the Ermakov-Lewis invariant.}" Unfortunately, the authors
miss the reference {\it Dynamical symmetries and the
Ermakov invariant}, by Haas and Goedert \cite{Haas1}. In this
paper, the Ermakov invariant is, apparently for the first time,
deduced as a consequence of a dynamical Noether symmetry. To make
this point clear, it is enough to compare equation (29) in the
paper by Haas and Goedert with  Proposition 1 and  equation (4.17)
in the paper by Moyo and Leach. Both equations present the
dynamical symmetry associated to the Ermakov invariant, for the
case of Lagrangian Ermakov systems, as the result from a
straightforward application of the converse of Noether's theorem.

We further notice that the Lagrangian formulation for Ermakov
systems in the referred publication \cite{Moyo} is by no means
new. This can be seen by comparing the potential functions from
equations (11) in the work by Haas and Goedert with the potential
given by equation (3.18) in the work by Moyo and Leach. This later
work also ignores the Hamiltonian descriptions for Ermakov systems
developed earlier in \cite{Cervero}, for the case of frequency
functions depending on time only, and in \cite{Haas2}, for the
case of frequency functions depending also on dynamical variables.

As a final remark, the work by Moyo and Leach also ignores the
papers \cite{Ray1}-\cite{Kaushal}, dedicated to the analysis of
uncoupled Ermakov systems in the light of Noether's theorem. These
works, however, do not deal with truly two-dimensional, coupled
Ermakov systems, as in the case of references \cite{Moyo, Haas1}.
Rather, these papers \cite{Ray1}-\cite{Kaushal} deal with Ermakov
systems in which one of the equations plays the principal role,
while the other, decoupled from the first, is treated as an
auxiliary equation. In these cases, the Lagrangian description is
effectively one-dimensional and the Ermakov invariant cannot be
obtained as a result from an associated dynamical Noether
symmetry.

\vskip 1cm \noindent{\bf Acknowledgements}\\  This work has been
supported by the Brazilian agency Conselho Nacional de
Desenvolvimento Cient\'{\i}fico e Tecnol\'ogico (CNPq).

 \end{document}